\documentclass[12pt]{article}

\usepackage{sbc-template}
\usepackage[caption=false]{subfig}
\usepackage{graphicx,url}
\usepackage[latin1]{inputenc}  

\usepackage{multirow} 
\usepackage{mdwmath}
\usepackage{framed}
\usepackage{tikz}
\usepackage{xcolor}
\usepackage{inputenc}  
\usepackage{amsmath,amssymb,array}
\usepackage{float} 
\usepackage{fancyhdr}
\usepackage{epstopdf}

\floatstyle{plain} 
\newfloat{task}{thp}{lst}
\floatname{task}{Task}
    
\title{Vis4DD: A visualization system that supports Data Quality Visual Assessment}

\author{João Marcelo Borovina Josko\inst{1}, João Eduardo Ferreira\inst{1}}
\address{Department of Computer Science \\ Institute of Mathematics and Statistics (IME) \\
  University of São Paulo (USP) \\ São Paulo - SP - Brazil \\
\email{\{jmbj,jef\}@ime.usp.br}
}

\begin{document} 

\maketitle


\begin{abstract}
Data quality assessment process is essential to ensure reliable analytical outcomes. This process depends on human supervision-driven approaches since it is impossible to determine a defect based only on data. Visualization systems belong to a class of supervised tools that can make data defect pattern visible. However, their considerable design knowledge encodings and implementations provide little support design to data quality visual assessment. 
To cover this gap, this work reports the design approach of $Vis4DD$ visualization system based on patterns of data defects structures and assessment tasks. An exploratory case study used this web-based system to explore which and how visual-interactive properties facilitate visual detection of data defect. 
\end{abstract}

\begin{quote} 
\end{quote}


\section{Introduction} \label{sec:int}

Data Quality Assessment process provides practical inputs to improve and keep data quality at levels required by analytical initiatives. Relevant computational models support such process, especially for data defects whose detection rules are more precise (e.g., Domain Constraint Violation \cite{Josko16b}). Such models are based on quantitative or constraint approaches that restrict the human role in interpreting their outcomes \cite{Dasu13}. 

On the other hand, data quality assessment process strongly depends on data context knowledge since it is impossible to confirm or refute a defect based only on data \cite{Dasu13}. The context specifies the structure of meaning and relationship between data and an environment (e.g., organization departments). Hence, human supervision is essential throughout this process. 

Visualization systems belong to a class of supervised approaches that combine computational capability with pattern-finding and semantic distinctions innate to human beings to permit data quality visual assessment. 


Much literature has encoded design knowledge regarding visualization systems, including perceptual-driven \cite{Ware04} perspectives. Related to data quality assessment, this knowledge has been encoded through certain implementations \cite{Chen15} or evaluation studies \cite{Marghescu07}. However, the analysis of this literature mostly reveals concerns about \textit{communicating} quality metrics measured on data with physical reference (e.g., a map) and little concern on how to permit \textit{visual comprehension and assessment} of data defect structures on abstract data (e.g., sales and billing).

To address this issue, this works introduces a web visualization system (named \textit{Visualization for Defect Detection} or $Vis4DD$) that supported an exploratory case study to identify which visual-interactive properties were more suitable for certain data defects structures on abstract data \cite{Josko16a}. Its design considered data defect structures, strategy patterns of visual assessment tasks and case study goals as inputs. 

The work reported here is organized as follows: Section~\ref{sec:domain} describes requirements and design issues related to $Vis4DD$ system, while Section~\ref{sec:arc} presents its components. Section~\ref{sec:dqa} conducts a comprehensive $Vis4DD$ walk-through and it briefly discusses certain case study findings. Section~\ref{sec:rw} outlines related works and Section~\ref{sec:con} concludes this work.


\section{Vis4DD Problem Domain, Requirements and Design} \label{sec:domain}

Data quality visual assessment denotes a nonlinear analytical process of comprehension of current data quality state mediated by visualization systems. Through interactive visual representations, data quality appraisers pursuit and correlate meanings (patterns and relationships) associated with a target defect structure until they integrate semantic evidences to confirm or refute it. Hence, absence of correspondence between a visual representation and this process goal prevents data quality appraisers from accomplishing their work. 

Visualization system design is manifold since there are different techniques composition that eventually may lead to an intended result. To offer a proper support to aforementioned problem domain, most $Vis4DD$ features were based on patterns of high-level tasks. These tasks denote cognitive strategies of visual inquiry in assessing data quality according to defect structures. 
 
The requirement analysis stage followed three steps that relied on a 6-year data quality analyst. The first step associated patterns with each data defect of case study interest according to their structure. The second step modelled and formalized high-level assessment tasks. For a complete task notation and formalization discussion, refer to \cite{Josko16a, josko2016c}. The last step analysed all modelled tasks characteristics to identify strategy patterns in regard to data simplification, space arrangement and visual abstraction. The case study goals added another set of requirements, including color scales, homogeneous visual representation appearance and log recording. 

Guided by the requirements analysis outcomes, the design stage followed three steps. The first decomposed the system into components (Section~\ref{sec:arc}), while the second step selected the most appropriate interactive techniques related to each strategy pattern. For instance, in case of space arrangement pattern we selected ordering, attribute arrangement and trellis. The last step followed the case study goals to select visualization techniques of different visual variables (e.g., position, hue, saturation, size, connection) and encoding types (e.g., point, line, proportionality, directed link).



\section{Vis4DD System Characteristics} \label{sec:arc}

\begin{figure}[!ht]
\center
\includegraphics [height=4cm]{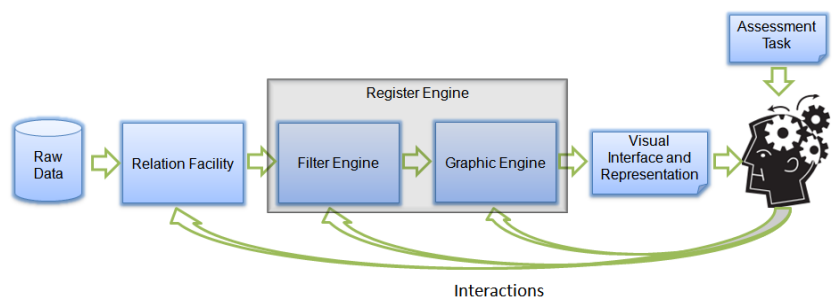}
\caption{$Vis4DD$ components communication (Source: Elaborated by the authors)}
\label{components}
\end{figure}

Figure~\ref{components} presents $Vis4DD$ architecture style and its components communication flow. These components are based on R language due to its analytic-driven features. $Vis4DD$ visual representations used Shiny framework to compose several visualization techniques, including parallel coordinates, radial graph, heat map, scatter plot matrix and tableplot. This framework provides an easy way to build web interactive solutions through a reactive programming model. Such model permits to control how (\textit{reactive conductors}) interface parameters (\textit{reactive sources}) changes elements of visual representations (\textit{reactive endpoint}). 

The \textit{Relation Facility} component enables managing (e.g. loading, discarding) any relation of interest in a R workspace. Relations must be first extracted from source databases as a formatted file to avoid interference in their operations and to provide a static data state for quality assessment. $Vis4DD$ provides different separators and quotes settings to load a formatted file. This operation keeps all original data values untouchable, but it executes certain structural checks (e.g., each line complies with file's header) and adjusts (e.g., convert numerical attribute into character when one of its value is not numerical).

The \textit{Filter Engine} selects data of interest according to multiple search criteria or pointing visual items. The multiple criteria denote a set of keywords for categorical attributes or range of values for quantitative attributes. The \textit{Graphic Engine} builds visual representations based on visualization technique, data characteristics and interactions parameters defined at visual interface. This component can handle all
data or selected data regions according to Filter Engine definition. The \textit{Register Engine} logs automatically all session interactions and their corresponding parameters. It is also in charge of taking visual representation shots when required by a data quality appraiser.


$Vis4DD$ implementation provides a rich set of visualization techniques displayed on independent visual scenes. Each scene allows certain interactions (e.g, geometric zooming, ordering, filtering, attribute arrangement, occlusion reduction) according to the visualization technique characteristics. Moreover, this implementation applies a segmented and unsegmented color scales (based on \textit{Hue, Saturation, Lightness} model) to ensure value distinctions on dense data spaces and quantitative data isomorphism, respectively.


\section{Data Quality Visual Assessment through Vis4DD} \label{sec:dqa}

\subsection{Walk-Through} 

$Vis4DD$ system starts working by loading the last saved R workspace and setting global parameters. In case of an empty workspace, all visualization techniques remain unavailable until the presence of any relation. 

In the early stage of data quality assessment, data quality appraisers may obtain an overall sense of all data and their patterns. They select an appropriate visualization technique to expose all data of a relation of interest. They proceed providing the corresponding target and reference attributes, and may also change default parameters of any interaction. For instance, some data quality appraiser may expose categories in different panels through trellis (e.g., Figure~\ref{atip2v_ok}). At the end of this setting procedure, data quality appraisers request the generation of the corresponding visual representation.

Interactions help data quality appraisers arrange data for comparison and correlation until they can isolate data regions potentially defective. In this stage, filtering and geometric zooming permit an easy and continuous refinement of data regions that are object of quality analysis. In case of strong suspicious, data quality appraisers can mark the defective data items (e.g., Figure~\ref{atip4v_ok}) and save the current visual representation. Otherwise, they can return to overall data view (by resetting interactions parameters) and recommences their analysis transitions until confirm or refute the presence of a data defect. At any time, a different visualization technique may be selected reusing the parameters already chosen.


\subsection{Case Study Summary} \label{sec:css}

Our exploratory case study used $Vis4DD$ to identify a set of relationships that exposes visual-interactive properties that permit visual assessment of different data defects. One of these data defects (atypical tuple) is outlined in this section. For a depth discussion of all data defects, refer to \cite{Josko16a, josko2016c}. 
 
\begin{figure}[!h]
\centering
\subfloat[{\scriptsize Atypical tuples ($2^{\underline{nd}}$ variant) detection through compacted frequency in hue in resolution of $10^7$ tuples}] {\includegraphics [height=5.7cm]{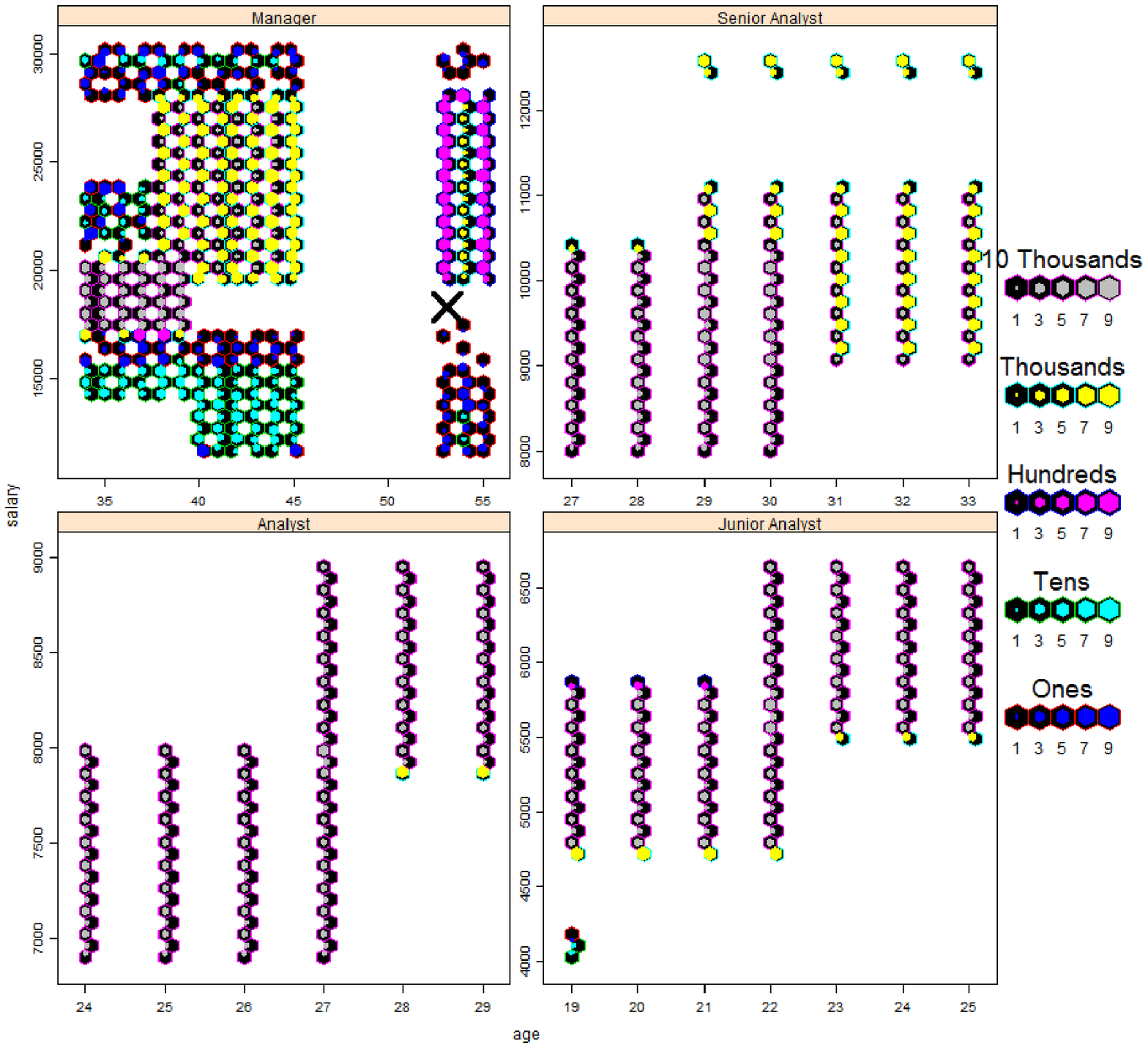} \label{atip2v_ok}} 
\subfloat[{\scriptsize Atypical tuples ($4^{\underline{th}}$ variant) detection through size proportional to average supported by image zooming in resolution of $10^6$ tuples}] {\includegraphics [height=5.7cm]{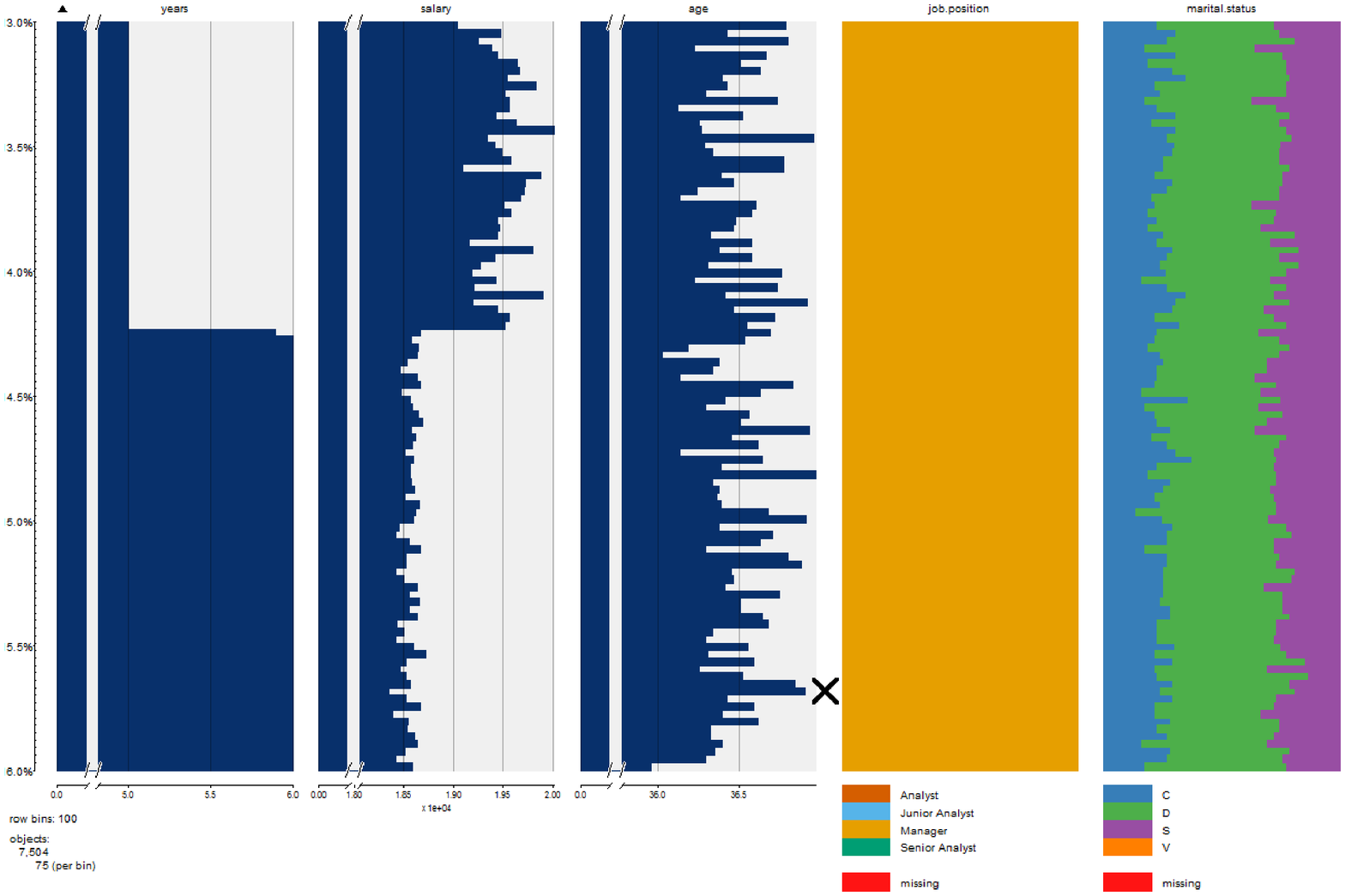} \label{atip4v_ok}}  
\centering
\caption{Portions of assessment scenes of Atypical Tuple variants (Source: \cite{Josko16a, josko2016c})}
\label{fig_result1}
\end{figure}

An atypical tuple deviates from the behavior of the remaining tuples of a relation for different reasons \cite{Josko16b}. Our case study considered four atypical variants. The $1^{\underline{st}}$ and $2^{\underline{nd}}$ variants denote $0.1\%$ and $1\%$ of defective values in an attribute, respectively. Most visual representations permitted their assessment, but position-based visualizations were outstanding. They made easy to perceive the structures of both variants, as the atypical ``manager salary" indicated in Figure \ref{atip2v_ok}. 

Position-based visualizations were also the best option to assess $3^{\underline{rd}}$ atypical value variant, although they required more interaction actions (e.g., filtering and point displacement). Such variant denotes atypical values interposed among data categories with certain superimposition.

The last variant ($4^{\underline{th}}$) denotes unusual combination of values considering multiples attributes. Due to its characteristics, only multidimensional visualizations permitted partial detection of atypical cases through intensive use of filter and zooming interactions. Figure~\ref{atip4v_ok} illustrates a $4^{\underline{th}}$ variant case involving ``years", ``salary" and ``age" attributes.



\section{Related Works} \label{sec:rw}
Knowledge concerning the design of visualization systems is encoded in different perspectives and depth levels. Due to the huge literature and space restrictions, this work only introduces implementation papers. For a broad discussion of such literature and its limitations in regard to data quality visual assessment, refer to \cite{Josko16a, josko2016c}.

Most implementation literature describes visualization systems based on \textit{Quality-Aware} approach to support Data Quality Assessment \cite{Chen15, Kandel12}. Such approach optimizes visualization techniques to communicate data quality metrics (extracted by computational resources) about a particular data defect. This sort of communication is useful for those data defects that require low-moderate human supervision (e.g., Domain Constraint Violation \cite{Josko16b}) or are visually imperceptible. 

However, these optimized visualizations do not consider visual properties according to data defect structure being assessed. Hence, this nonalignment obstructs extraction and comprehension of its meanings due to the distraction effect \cite{Ware04}.

On the other hand, few literature describes visualization systems that support extensive use of visual exploratory analysis of meanings to determine defective data \cite{Tennekes13, Fuhring07}. The supervised nature of this visual approach (named \textit{Visual Diagnosis-Driven}) is basis for those defects whose analysis strongly depends on human supervision and contributions from computational resources (when available) are restricted. However, it is unclear \textit{if} and \textit{how} these aforementioned systems considered data defect structures, visual assessment tasks or backing of data quality experts to guide their design choices. Our analysis revealed a lack of proper alignment between chosen visual-interactive properties and data defects intended of assessment.


\section{Conclusions} \label{sec:con}
This work reports the design approach and components of $Vis4DD$ visualization system that supports quality visual assessment on abstract data. Its characteristics enabled the analysis of which and how different visual-interactive properties facilitated (or not) the perception and comprehension of meanings in regard to data defect structures  that requires high level of human supervision. Nevertheless, $Vis4DD$ neither addresses multiple coordinated views nor offers computational approaches (e.g. data mining methods) for data defects without visual evidence. As future works, it is intended to provide features to associate quality assessment outcomes to data (annotation), extract data straight from relational databases and apply creativity techniques to a broader range of data quality analysts to stimulate new ideas.


\section{Acknowledgments}
This work has been supported by CNPq (Brazilian National Research Council) grant number 141647/2011-6 and FAPESP (S\~ao Paulo State Research Foundation) grant number 2015/01587-0.


\bibliographystyle{sbc}
\bibliography{joskoBib}

\end{document}